\documentstyle[aps,epsfig]{revtex}

\begin{document}
\draft

\title{The light $\sigma$-meson}
\author{V.V. Anisovich and V.A. Nikonov }
\address{St.Petersburg Nuclear Physics Institute, Gatchina, 188350,
Russia}
\date{\today}
\maketitle

\begin{abstract}
In the framework of the dispersion relation
$N/D$-approach, we restore the low--energy $\pi\pi$
$(IJ^{PC}=00^{++})$-wave amplitude sewing it with the previously
obtained $K$-matrix solution for the region 450--1900 MeV.
The restored $N/D$-amplitude has a pole on the second sheet of the
complex-$s$ plane near the $\pi\pi$ threshold, that is the light
$\sigma$-meson.
\end{abstract}

\pacs{12.39.Mk, 12.38.-t, 14.40.-n}

At present the understanding of scalar meson is one of the
key problems for the Strong-QCD physics. The $\pi\pi$ low-mass data
provide indications on the existence of a low-mass $\sigma$-meson.
This state is beyond $q\bar q $ and gluonium systematics, which
makes it necessary  to confirm its existence as well as to
 study the possible mechanisms of its formation.

Experimental data on meson spectra accumulated by the Crystal
Barrel Collaboration \cite{CrBar}, GAMS \cite{GAMS} and BNL \cite{BNL}
groups provided a good basis for setting up the $q\bar q$/gluonium
classification of the light scalars. For the
$(IJ^{PC}=00^{++})$-wave,
the combined $K$-matrix analysis of the reactions $\pi\pi \to \pi\pi$,
$K\bar K$, $\eta\eta$, $\eta\eta'$, $\pi\pi\pi\pi$ has been carried out
over the mass range 450--1900 MeV  \cite{APS,AKPSS}, then the $K$-matrix
analysis was extended to the waves $\frac12 0^{+}$ \cite{AlexSar}
and $10^{+}$ \cite{AKPSS}, thus making it possible to establish the
$q\bar q$ systematics of scalars for $1^3P_0 q\bar q$ and $2^3P_0 q\bar
q$ multiplets.

The advantage of the $K$-matrix representation
is that it allows us  not only to determine the locations
and partial widths of resonances but also to study characteristics
of corresponding states with switched off decay channels: these
"primary states", or "bare states" (i.e. states without a cloud of
mesons produced by decay processes) are suitable objects
for the $q\bar q$/gluonium classification
(see \cite{AAS} for the details). The decay processes in the
scalar/isoscalar sector cause a strong mixing which destroys the $q
\bar q$/gluonium classification.  Another important effect
generated by transitions $(q\bar q)_1\to real\; mesons \to (q\bar
q)_2$ is an
accumulation of widths of  neighbouring resonances by one of them,
that results in appearance of a broad state.

According to \cite{APS,AKPSS}, five bare scalar/isoscalar
states are located
in the region 700-1800 MeV:
$f_0^{bare}(720\pm 100)$,  $f_0^{bare}(1230\pm 50)$,
$f_0^{bare}(1260\pm 30)$,  $f_0^{bare}(1600\pm 50)$  and
$f_0^{bare}(1800\pm 30)$.
Four of them are members of the $q\bar q$ nonets
$1^3P_0 q\bar q$ and $2^3P_0 q\bar q$ while one state,
$f_0^{bare}(1260\pm 30)$ or $f_0^{bare}(1600\pm 50)$,
is the lightest glueball; results of the lattice calculations
\cite{lattice} tell us
that it is
$f_0^{bare}(1600\pm 50)$.
 After the mixing originated from
the decay processes, the primary, or bare states
are transformed into a set of resonances:
$f_0(980)$, $f_0(1370)$, $f_0(1500)$,
$f_0(1530\; ^{+90}_{-250})$
and $f_0(1750)$. The state $f_0(1530\; ^{+90}_{-250})$
is rather broad, its
large width is due to the accumulation of widths of
neighbouring resonances: the gluonium and $q\bar
q$ states are strongly mixed because
the transition $gluonium \to q\bar q$
is not suppressed
in terms of the rules of $1/N$
expansion \cite{t'Hooft}.
The gluonium component is shared between the the broad
state $f_0(1530\; ^{+90}_{-250})$
and the scalars $f_0(1370)$ and $f_0(1500)$.

An important result of the article \cite{APS,AKPSS}  is
that the $K$-matrix $00^{++}$-amplitude  has no
pole singularities in the region 500--800 MeV. Here the
$\pi\pi$-scattering phase $\delta^0_0$ increases smoothly
reaching 90$^\circ$ at 800--900 MeV. A straightforward
explanation of such a behaviour of $\delta^0_0$ might consist in the
presence of a broad resonance, with a
mass  about 600--900 MeV and width $\Gamma \sim 500$ MeV
(for example, see Refs. \cite{M,O} and references therein).
However, according to the $K$-matrix solution \cite{APS,AKPSS},
the $00^{++}$-amplitude does not contain  pole singularities
on the second sheet of the complex-$M_{\pi\pi}$ plane
inside the interval $450 \leq Re\; M_{\pi\pi}\leq 900 $
MeV: the $K$-matrix amplitude has a  low-mass pole
only, which is located on the  second sheet either near the
$\pi\pi$ threshold or even below it. In \cite{APS,AKPSS}, the
presence of this pole was not emphasized, for the left-hand cut,
which is important for the reconstruction of analytic structure
of the low-energy partial amplitude, was taken
into account only indirectly; a proper way for the description of
the low-mass amplitude must be the dispersion relation
representation.

In this
paper the $\pi\pi$-scattering $N/D$-amplitude
is reconstructed
in the region of small $M_{\pi\pi}$ being sewed
with the $K$-matrix solution
\cite{APS,AKPSS} found for $M_{\pi\pi}\sim 450-1950 $ MeV.
More specifically, using the data for $\delta^0_0$,
we construct the $N/D$ amplitude below 900
MeV sewing it with the $K$-matrix amplitude, bearing in mind to make a
continuation of the amplitude
into the region $s=M^2_{\pi\pi}\sim 0$. With this sewing we
strictly follow the results obtained for the $K$-matrix  amplitude in the
region 450-900 MeV, that is, the region where we can trust
the $K$-matrix representation of the amplitude. Recall that
the $K$-matrix
representation allows us to restore correctly the analytic structure of
the amplitude in the region $s > 0$ (by taking into account
threshold and pole singularities)
but not the left-hand singularities at $s\leq 0$ (where
singularities are related
to forces). Therefore, being cautious, we cannot be quite confident
of the $K$-matrix results below $\pi\pi$ threshold.

The dispersion relation amplitude is reconstructed
with the method of
approximation of the left-hand cut suggested in \cite{AKMS}. The
found $N/D$-amplitude provides us with a good description of
$\delta^0_0$ from the threshold to 900 MeV, that includes the region
where $\delta^0_0 \sim 90^\circ$. This amplitude does not
have a pole in the region 500--900 MeV; instead, the pole is located
near the $\pi\pi$ threshold.
This pole corresponds to the light $\sigma$-meson.

\section{Dispersion relation representation for the $\pi\pi$
scattering amplitude below 900 MeV}

The partial pion-pion scattering amplitude being
a function of the invariant energy squared
$s=M_{\pi\pi}^2$ can be
represented as a ratio $N(s)/D(s)$ where $N(s)$ has a left-hand cut
which is due to  "forces" (interactions due to $t$- and
$u$-channel exchanges), while $D(s)$ is determined by the
rescatterings in the $s$-channel. $D(s)$ is given by the
dispersion integral along the right-hand cut in the complex-$s$ plane:
\begin{equation}
A(s)=\frac{N(s)}{D(s)}\; , \;\;\;D(s)=1-\int
\limits_{4\mu^2_\pi}^\infty \frac
{ds'}{\pi} \frac{\rho(s')N(s')}{s'-s-i0}\; .
\end{equation}
Here $\rho(s)$ is the invariant $\pi\pi$ phase space,
$\rho(s)=(16\pi)^{-1}
\sqrt{(s-4\mu^2_{\pi})/s}$. It was supposed in (1) that $D(s) \to 1$
with $s\to \infty$ and CDD-poles are absent
(a detailed presention of the
$N/D$-method can be found in \cite{Chew}).

The $N$-function can be written as an integral along the left-hand
cut as follows:
\begin{equation}
N(s)=\int
\limits_{-\infty}^{s_L}  \frac{ds'}{\pi}\frac{L(s')}{s'-s}\; ,
\end{equation}
where the value $s_L$ marks the beginning of the
left-hand cut. For example,
for the one-meson exchange contribution $g^2/(m^2 -t)$, the
left-hand cut
starts at $s_L=4\mu_\pi^2-m^2$, and at this point the
$N$-function has a logarithmic singularity; for
the two-pion exchange $s_L=0$.

Below we deal with the amplitude $a(s)$ which is defined as follows:
\begin{equation}
a(s)= \frac {N(s)}{8\pi \sqrt{s}\left (1-P\int
\limits_{4\mu_\pi^2}^\infty
\frac{ds'}{\pi}\frac{\rho(s')N(s')}{s'-s}\right ) }\; .
\end{equation}

The amplitude $a(s)$  is related to the scattering phase shift:
$a(s)\sqrt{s/4-\mu_\pi^2} = \tan
\delta^0_0$.  In Eq. (3) the threshold singularity is singled out
explicitly, so the function $a(s)$  contains only the left-hand cut
together with poles corresponding to zeros
of the denominator of the
right-hand side in (3) which follow from:
$1=P\int\limits _{4\mu_\pi^2}^\infty
(ds'/\pi)\cdot \rho(s')N(s')/(s'-s) $. The pole of $a(s)$
at $s>4\mu_\pi^2$ corresponds to the phase shift
value $\delta^0_0 = 90^\circ$. The phase of the $\pi\pi$
scattering  reaches the value $\delta^0_0 = 90^\circ$ at $\sqrt{s}=
M_{90}\simeq
850$ MeV. Because of that, the amplitude $a(s)$ may be represented in
the form:
\begin{equation}
a(s)=\int\limits_{-\infty}^{s_L}  \frac{ds'}{\pi}\frac{\alpha(s')}{s'-s}+
\frac{C}{s-M^2_{90}}+D.
\end{equation}
To reconstruct the low-mass amplitude the parameters
$D,C,M_{90}$ and $\alpha(s)$ have been determined by fitting to the
experimental data.  In the fit we have used a method, which has been
established in the analysis of the low-energy nucleon-nucleon amplitudes
\cite{AKMS}.  Namely, the integral in the right-hand side of (4) has
been replaced by the sum
\begin{equation}
\int\limits_{-\infty}^{s_L}
\frac{ds'}{\pi}\frac{\alpha(s')}{s'-s} \to \sum_{n} \frac{\alpha_n}{s_n
-s}
\end{equation}
with $ -\infty < s_n \leq s_L$.

In the fit to the data for $\delta^0_0$ at $\sqrt s
\leq 950$ MeV, see Fig. 1a, we impose the following constraints on the
$N/D$-solution in order to sew it to the $K$-matrix
amplitude found previously at $\sqrt s \sim $450--1950 MeV
\cite{APS,AKPSS}:

(i) The $N/D$-solution curve for $a(s)$, see Fig. 1b, should be inside
the corridor determined by the
$K$-matrix solution at 450 MeV$\leq \sqrt s \leq $950 MeV: the corridor
in Fig. 1b is shown by the error bars for the $K$-matrix solution
points.

(ii) The $N/D$-amplitude should be analytical (not having pole
singularities) in the following complex-$s$ region on the second sheet:
$ 0.25$ GeV$^2\leq \mbox {Re} \; s\leq 0.8$ GeV$^2$ and $ 0 \leq \mbox
{Im} \; s\leq 0.6$ GeV$^2$.

The description of data within the $N/D$-solution,
which uses six terms in the sum (5), is demonstrated on Fig. 1a.
The parameters of the solution are as follows (all values are in
$\mu_\pi$ units):
\begin{equation}
M_{90}=6.228, \qquad C=-13.64,\qquad D=0.316 \; ,
\end{equation}
$$
(\alpha_n,\;s_n)= (2.23,\; - 9.56),\; ( 2.21,\; -10.16),\;), \;
( 2.19,\; -10.76),
$$
$$
(0.247,\; -32),\; (0.246,\; -36),\; (0.245, \; -40) .
$$
 The scattering length found for this solution is equal to
$a^0_0=0.22\; \mu_\pi^{-1}$
(experiment gives $a^0_0=0.26\pm 0.05\; \mu_\pi^{-1}$  \cite{a0}),
the Adler zero
is at $ s=0.12 \; \mu_\pi^2$. The
$N/D$-amplitude is sewed with the $K$-matrix
amplitude of Refs. \cite{APS,AKPSS}, and figure 1b demonstrates the
level of coincidence for the amplitudes  $a(s)$ for both solutions
(the values of $a(s)$ which correspond to the $K$-matrix amplitude are
shown with error bars determined in \cite{APS,AKPSS}).

The dispersion relation solution has correct analytic structure at
the region $|s|<1$ GeV$^2$.  The amplitude has no
poles on the first sheet of the complex-$s$ plane;
the left-hand cut of the $N$-function after
the replacement given by Eq. (5) is transformed into a set of poles
on the  negative piece of
the real $s$-axis: six poles of the amplitude (at $s/\mu_{\pi}^2=
-5.2,\; -9.6,\; -10.4,\; -31.6,\; -36.0,\; -40.0$)
represent the left-hand singularity of $N(s)$.
On the second sheet (under the $\pi\pi$-cut) the amplitude has two
poles:  at $s\simeq (4-i14)\mu^2_{\pi}$  and $s\simeq
(70-i34)\mu^2_{\pi}$ (see Fig. 2). The second pole, at
$s=(70-i34)\mu^2_{\pi}$, is located beyond the region under
consideration for which $|s|<1$ GeV$^2$
(nevertheless, let us stress that the $K$-matrix amplitude
\cite{APS,AKPSS} has a set of poles just in the region
of the second pole of the $N/D$-amplitude).
The pole near the threshold, at
\begin{equation}
s\simeq (4-i14)\mu^2_{\pi} \; ,
\end{equation}
is what we discuss.
The $N/D$-amplitude has no poles at $\mbox {Re} \sqrt
s \sim 600-900$ MeV despite the phase shift
$\delta^0_0$ reaches $90^\circ$ here.

In the solution discussed above, Eq. (6), the left-hand singularity is
described by six poles. With this number of poles, the solution
is weakly depending on their change: for example, the five-pole solution
with $a^0_0=0.22\; \mu_\pi^{-1}$ gives practically the same result
at $\mbox {Re} \; s > 0$ as the six-pole one.

The data do not fix the $N/D$-amplitude rigidly. The position of the
low-mass pole can be easily varied in the region
$\mbox {Re}\; s \sim (0 - 4)\mu_{\pi}^2$, and there are subsequent
variations of
the scattering length in the interval $a^0_0 \sim (0.21  - 0.28 )
\mu^{-1}_\mu $ and Adler zero at $s\sim
(0-1)\mu_{\pi}^2$.
Ambiguities in fixing the $\sigma$-pole and Adler
zero positions are mainly due to comparatively large error bars
in the measured $\delta^0_0$ near threshold, at $\sqrt s < 350$ MeV.

Let us stress that the way of reconstruction of the dispersion relation
amplitude used here differs from the mainstream attempts to
determine the $N/D$-amplitude. In the classical $N/D$ procedure,
that is the bootstrap one, the pion--pion amplitude is to be determined
by analyticity, unitarity and crossing symmetry. This means a unique
determination of the left-hand cut by the crossed channels.
However the bootstrap procedure is not carried out till now; the
problems which faces the nowadays bootstrap
program are discussed in Ref. \cite{VV} and references therein.
Nevertheless, one can try to saturate the left-hand cut by
known resonances in the crossing channels. Usually one supposes
that the dominant contribution to the left-hand cut comes from
the $\rho$-meson exchange supplemented by $f_2(1275)$ and
$\sigma$ exchanges. Within this scheme the low-energy amplitude is
restored, being controlled by the available experimental data.

In the scheme used here the amplitude
in the physical region at 450-1950 MeV is supposed to be known from
the $K$-matrix analysis, and then a continuation of the amplitude
is performed from 450-900 MeV to the region of smaller
masses; the continuation is restricted by the data.  As a result, we
restore the pole near the threshold (the low-mass $\sigma$-meson) and
the left-hand cut (although with less accuracy, actually on a
qualitative level).

In the approaches which take into account
the left-hand cut as a contribution of certain
meson exchanges, the following locations of the low-mass pole were
obtained:\\
(i) dispersion relation approach, $s \simeq
(0.2-i22.5)\mu_\pi^2 $ \cite{Basdevant}, \\
(ii) meson exchange models,
$s \simeq (3.0-i17.8)\mu_\pi^2 $ \cite{Zinn}, $s \simeq
(0.5-i13.2)\mu_\pi^2 $ \cite{Bugg}, $s \simeq (2.9-i11.8)\mu_\pi^2 $
\cite{Speth}, \\
(iii) linear $\sigma$-model, $s \simeq
(2.0-i15.5)\mu_\pi^2 $ \cite{Achasov}.

However, in \cite{800P,800E,800A,800Ishida,Roos,Loch}, the pole
positions were found in the region of higher $s$, at $s> 7
\mu_\pi^2$, that reflects ambiguities of the approaches which
treat the left-hand cut as a known quantity.

\section{Conclusion}

On the basis of the dispersion relation $N/D$-representation, we
have continued the $K$-matrix $00^{++}$ amplitude found previously for
$\sqrt s = M_{\pi\pi} \sim 450 -1950$ MeV
\cite{APS,AKPSS} to the $\pi\pi$ threshold region, $s\sim
(0-4\mu^2_{\pi})$; the  continuation procedure has been corrected by
the low-energy data. The amplitude found in this way  has a pole
near the $\pi\pi$ threshold, at $Re \; s \sim (0- 4\mu^2_{\pi})$; this
pole corresponds to the light $\sigma$-meson.  This result is in a
qualitative agreement with that of Refs.
\cite{Basdevant,Zinn,Bugg,Speth,Achasov}
where the analysis of the $00^{++}$ amplitude was
performed by modelling the left-hand cut contribution.

With the results for
the $K$-matrix analysis of Refs. \cite{APS,AKPSS}, one has
six scalar/isoscalar states in the region below $1800$ MeV. Five
of them are descendants of
the $q\bar q$ states ($1^3P_0q\bar q$ and $2^3P_0q\bar q$)
and gluonium. They are $f_0(980)$, $f_0(1370)$,  $f_0(1500)$,
$f_0(1530\; ^{+30}_{-250})$ and $f_0(1750)$. Three
states, $f_0(1370)$,  $f_0(1500)$
and $f_0(1530\; ^{+30}_{-250})$,
shared the gluonium component.
There are arguments (see Ref. \cite{AAS} for details)
that the broad state
$f_0(1530\; ^{+30}_{-250})$ is a descendant of the lightest scalar
gluonium which mass, according to lattice calculations
\cite{lattice}, is in the region 1500--1700 MeV;  an appearance of the
broad state $f_0(1530\; ^{+30}_{-250})$ is due to the specific
effect of accumulation of widths of the overlapping resonances
\cite{locking}.
So, we conclude that the light $\sigma$-meson is
beyond both the $q\bar q$ and gluonium systematics.

\bigskip

We thank A.V. Anisovich, Ya.I. Azimov, D.V. Bugg, L.G. Dakhno,
H.R. Petry, A.V. Sarantsev, V.V. Vereshagin  and A.V. Yung for
fruitful discussions.
The article is supported by the RFBR grant N 98-02-17236.

\begin{figure}
\centerline{\epsfig{file=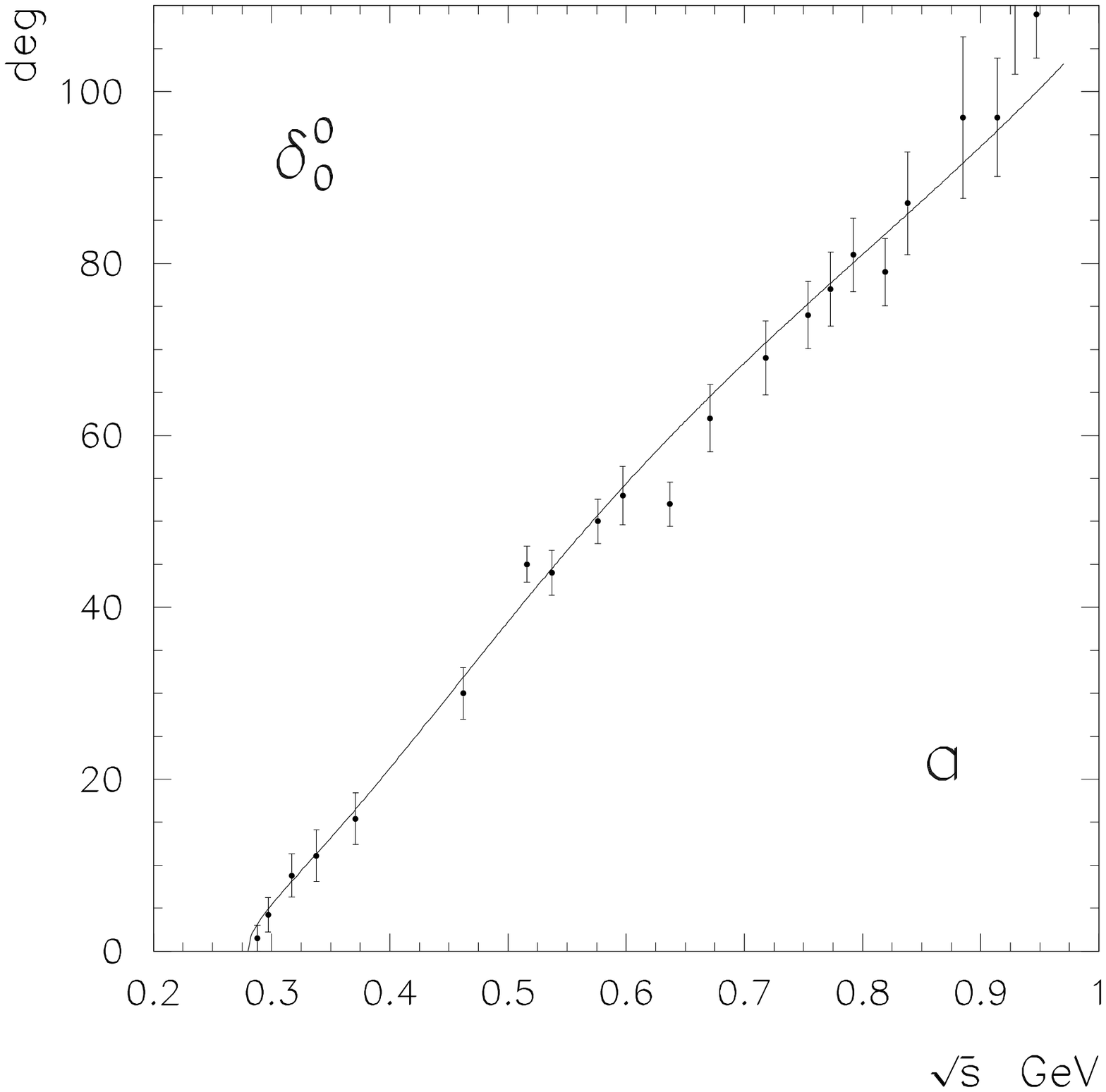,width=7.5cm}\hspace{1cm}
            \epsfig{file=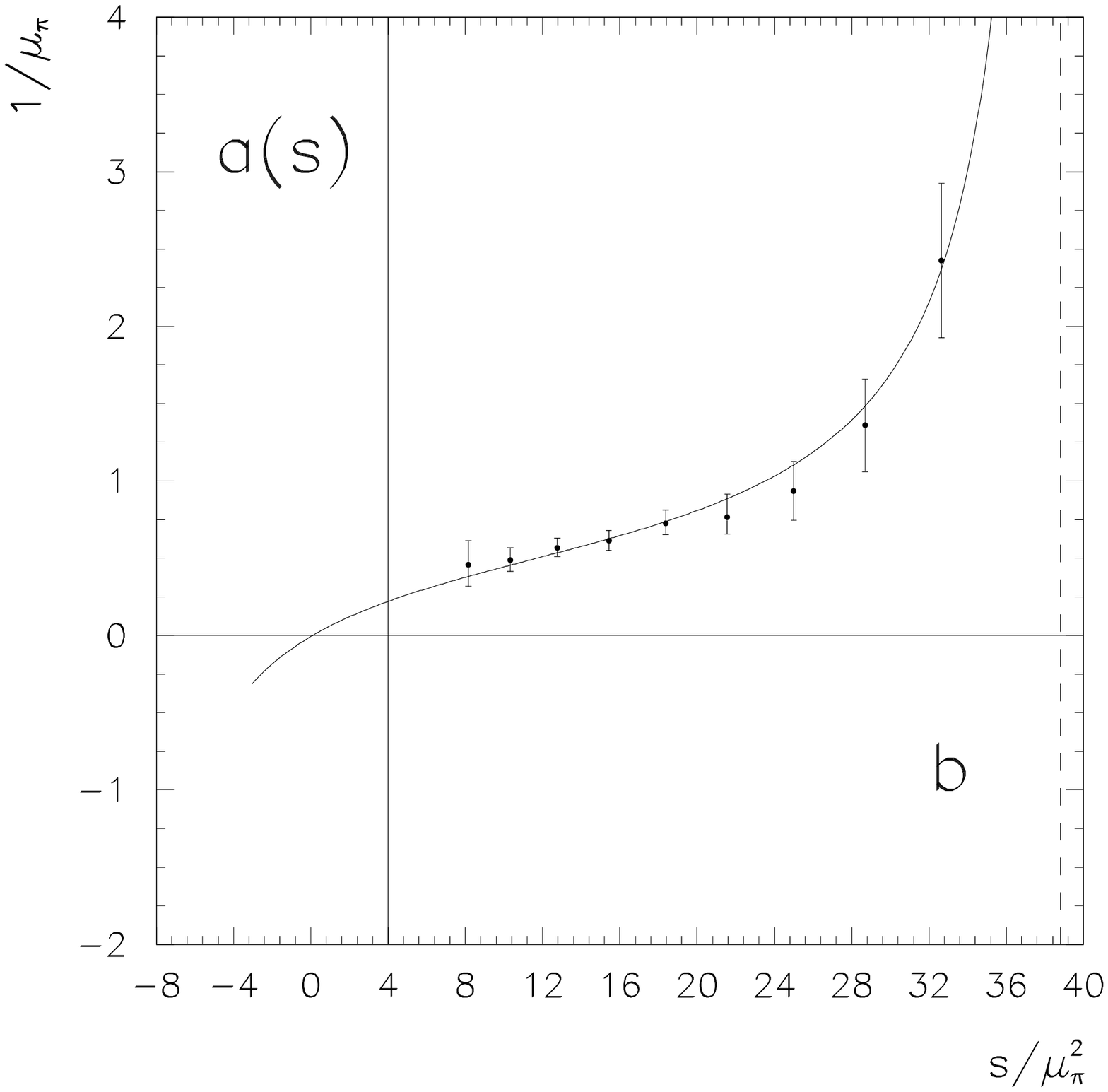,width=7.5cm}}
\caption{a) Fit to the data on $\delta^0_0$ by using the
$N/D$-amplitude.  b) Amplitude $a(s)$ in the $N/D$--solution
(solid curve) and the $K$-matrix approach [4,6] (points with
error bars). }
\end{figure}

\begin{figure}
\vspace{3cm}
\centerline{\epsfig{file=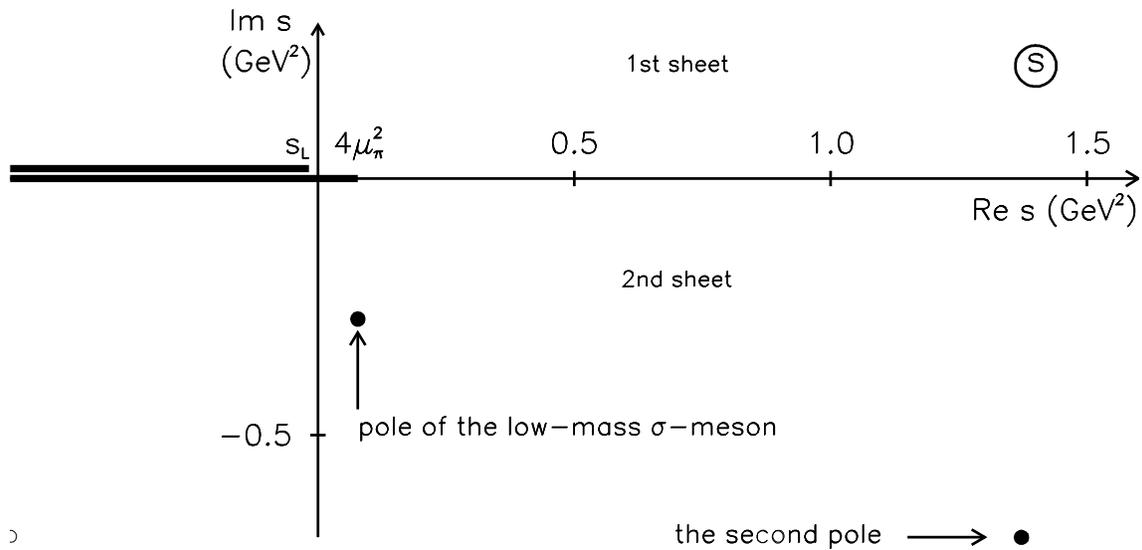,width=15cm}}
\caption{Complex-s plane and singularities of the $N/D$-amplitude}
\end{figure}

\end{document}